\documentclass[%
superscriptaddress,
 amsmath,amssymb,
 aps,
 prstab, twocolumn,
]{revtex4-1}

\usepackage{tabularx}
\usepackage{textcomp}
\usepackage{graphicx}%
\usepackage[dvipsnames]{xcolor}

\usepackage{amsmath}
\usepackage{amssymb}
\usepackage{dcolumn}
\usepackage{graphicx}
\usepackage{longtable}
\usepackage{bm}
\usepackage{color}
\usepackage{epsfig}
\usepackage[colorlinks, citecolor=blue, urlcolor=black, bookmarks=false, linkcolor=blue, hypertexnames=true]{hyperref}

\begin{document}

\title{Pressure-induced creation and annihilation of Weyl points in $T_d$- and 1$T''$-Mo$_{0.5}$W$_{0.5}$Te$_2$}

\author{Bishnu Karki}
\thanks{These authors contributed equally}
\affiliation{Central Department of Physics, Tribhuvan University, Kirtipur, 44613, Kathmandu, Nepal}
\affiliation{Condensed Matter Physics Research Center, Butwal-11, Rupandehi, Nepal}

\author{Bishnu Prasad Belbase}
\thanks{These authors contributed equally}
\affiliation{Central Department of Physics, Tribhuvan University, Kirtipur, 44613, Kathmandu, Nepal}
\affiliation{Condensed Matter Physics Research Center, Butwal-11, Rupandehi, Nepal}

\author{Gang Bahadur Acharya}
\affiliation{Central Department of Physics, Tribhuvan University, Kirtipur, 44613, Kathmandu, Nepal}
\affiliation{Institute for Theoretical Solid State Physics, IFW Dresden, Helmholtzstr. 20, 01069 Dresden, Germany}

\author{Sobhit Singh}
\email{sobhit.singh@rutgers.edu}
\affiliation{Department of Physics and Astronomy, Rutgers University, Piscataway, New Jersey 08854, USA}

\author{Madhav Prasad Ghimire}
\email{madhav.ghimire@cdp.tu.edu.np}
\affiliation{Central Department of Physics, Tribhuvan University, Kirtipur, 44613, Kathmandu, Nepal}
\affiliation{Condensed Matter Physics Research Center, Butwal-11, Rupandehi, Nepal}
\affiliation{Institute for Theoretical Solid State Physics, IFW Dresden, Helmholtzstr. 20, 01069 Dresden, Germany}


\begin{abstract}
By means of first-principles density-functional theory calculations, we investigate the role of hydrostatic pressure on the electronic structure of $T_d$ ($Pmn2_1$) and 1$T''$ ($Pm$) phases of Weyl semimetal Mo$_{0.5}$W$_{0.5}$Te$_2$, which is a promising material for phase-change memory technology and superconductivity. We particularly focus on changes occurring in the distribution of the gapless Weyl points (WPs) within 0 to 45 GPa pressure range. We further investigate the structural phase transition and lattice dynamics of the $T_d$ and 1$T''$ phases within the aforementioned pressure range. 
Our calculations suggest that both the $T_d$ and 1$T''$ phases of Mo$_{0.5}$W$_{0.5}$Te$_2$ host four WPs in their full Brillouin zone at zero pressure.
The total number of WPs increases to 44 (36) with increasing pressure {\it via} pair creation up to 20 (15) GPa for the $T_d$ (1$T''$) phase, and beyond this pressure pair annihilation of WPs starts occurring leaving only 16 WPs at 45 GPa in both phases. 
The enthalpy versus pressure data reveal that the 1$T''$ phase is more favorable below the critical pressure of 7.5 GPa, however, beyond this critical pressure the $T_d$ phase becomes enthalpically favorable. We also provide the calculated x-ray diffraction spectra along with the calculated Raman- and infrared-active phonon frequencies to facilitate the experimental identification of the studied phases.

\end{abstract}

\maketitle

\section{Introduction}
The study of topological materials is of high interest at present due to their potential applications in the emerging technology~\cite{RevModPhys.82.3045, RevModPhys.90.015001,doi:10.1146/annurev-conmatphys-031016-025458, manna2018heusler, weng2016topological,gao2019topological,burkov2018weyl,
hasan2017discovery}. 
Weyl semimetals (WSMs), one class of topological materials, seem to drag special attention since their experimental realization in 2015~\cite{lv2015observation,lv2015experimental, huang2015weyl, xu2015experimental, PhysRevB.92.241108, PhysRevLett.115.217601}.   
Breaking of either time-reversal symmetry or spatial inversion-symmetry, or both, in WSMs results in a particular electronic band structure possessing crossings of the nondegenerate valance and conduction bands near the Fermi level forming gapless Weyl points (WPs) and Weyl cones~\cite{murakami2007phase}. The low-energy electronic excitations near these WPs behave as the massless Weyl fermions~\cite{jia2016weyl, Weyl1929elektron, PhysRevB.83.205101, grassano2018validity, ruan2016symmetry, Soluyanov_Nat2015, singhprb2016, singhprm2018, Winkler_2019}. 
WSMs are interesting due to their exotic properties such as WPs acting as the sources and sinks of the Berry curvature in momentum space, existence of open Fermi-arc states connecting two opposite WPs,
extremely large magnetoresistance~\cite{FCChenPRB2016, QLPei_PRB2017,
Thirupathaiah_PRB2017,  SangyunLee_SciRep2018}, and 
various quantum Hall phenomena~\cite{xu2015discovery,xu2016observation, PhysRevLett.116.066802, ATamai_PRX2016, autesprl2016, zhangprb2017, ghimire2019creating, SinghPRL2020}.
Moreover, based on the tilting of the band crossings near the Fermi level two types of WSMs are reported: (i) type-I WSMs that preserve the Lorentz invariance and their Fermi surface shrinks to zero when the Fermi energy (E$_F$) is set at the energy of the WPs, 
and (ii) type-II WSMs that violate the Lorentz invariance due to peculiar tilting of the Weyl cone such that WPs occur at the touching points of an electron and hole pockets and the Fermi surface never shrinks to absolute zero when E$_F$ is set at the energy of the WPs~\cite{doi:10.1146/annurev-conmatphys-031016-025458,grassano2018validity,Soluyanov_Nat2015,PhysRevLett.117.066402,chang2016strongly,PhysRevB.93.201101}.

WSMs belonging to the transition metal dichalcogenides (TMDs) family are special, mainly because the \textit{s}, \textit{p}, and \textit{d} orbitals in these systems hybridize to form bands near the Fermi level, which often yield various fascinating properties such as distinct quantum phase transitions between different structures, thermal and optical properties, topological domain walls, different kinds of Hall effects, and superconductivity~\cite{sipos2008mott,qian2014quantum,morris1972superconductivity, TakahashiPRB2017, qi2016superconductivity, chenprb2018, huang2019polar}. 
Notably, dome-shaped superconducting behaviour is observed in MoTe$_2$ with a transition temperature of 0.01 K that enhances to 8.2 K at 11.7 GPa pressure~\cite{qi2016superconductivity, HeikesPRM2018,PhysRevB.5.895,PhysRevB.89.014506,PhysRevB.90.045130}.  
MoTe$_2$ and WTe$_2$ are among the first reported WSMs with four WPs in their momentum space resulting due to the broken inversion symmetry~\cite{soluyanov2015type,li2018spin,bruno2016observation, wu2016observation,li2017evidence,lv2017experimental,huang2016spectroscopic, deng2016experimental,sakano2017observation, sun2015prediction,wang2016mote,jiang2017signature, liang2016electronic}. 
The application of external pressure and strain has been reported to play a vital role in tuning the WSM phase in MoTe$_2$ and WTe$_2$ \cite{YanSun_PRB2015, DissanayakeNPJ2019, Aryal_PRB2019, NXu_PRL2018}.

Substitution of W by Mo in WTe$_2$, forming a polymorphic structure of Mo$_x$W$_{1-x}$Te$_2$, is reported to control the structural phase transition, 
transport properties, thermal conductivity, Weyl phase, and superconductivity in this system~\cite{Lv2017, YanAPL2017, oliver2017structural,RhodesNL2017,AslanNL2018,SchneelochPRB2020, marchenkov2019electronic, TaoPRB2019}. 
Depending on the magnitude of the applied pressure on Mo$_x$W$_{1-x}$Te$_2$ ($x$ = 0.9, 0.6, and 0.25), the  superconducting transition temperature can be tuned~\cite{dahal2020tunable}.
Moreover, tunable WSM phase and Fermi-arc states can be realized in Mo$_x$W$_{1-x}$Te$_2$ ($x$ $\leq$ 0.5) by varying $x$~\cite{belopolski2016Fermi,chang2016prediction,belopolski2016discovery}. 
By means of the  temperature-dependent elastic neutron scattering measurements, Schneeloch {\it et al.}~\cite{SchneelochPRB2020} recently investigated the monoclinic 1$T'$ to orthorhombic $T_d$ structural phase transition in Mo$_{1-x}$W$_{x}$Te$_2$ as a function of the W substitution and temperature. They observed that 1$T'$ -- $T_d$ phase transition is accompanied by an intermediate pseudo-orthorhombic phase $T_d^{*}$, which was first reported by Tao {\it et al.}~\cite{TaoPRB2019} for pure MoTe$_2$. Notably, the intermediate $T_d^{*}$ phase exists only up to $x = 0.34$ concentration and vanishes for $x>0.34$. Their results further suggest that at room temperature Mo$_{0.5}$W$_{0.5}$Te$_2$ composition favors the $T_d$ phase~\cite{SchneelochPRB2020}. Marchenkov {\it et al.}~\cite{marchenkov2019electronic} experimentally studied the transport and optical properties of a single crystal Mo$_{0.5}$W$_{0.5}$Te$_2$. Their temperature-dependent resistivity data reveal the metallic nature of the system. Li {\it et al.} recently performed dimensionality-dependent transport measurements on a special Mo$_{0.5}$W$_{0.5}$Te$_2$ sample having a thickness gradient across 2D and 3D regimes and reported strong evidences that this material is a type-II WSM~\cite{PhysRevB.104.085423}.

The experimental observation of pressure-controlled superconductivity in Mo$_{0.5}$W$_{0.5}$Te$_2$~\cite{dahal2020tunable} calls for a systematic investigation of the pressure-induced changes occurring in the crystal structure, electronic structure, and WSM phase of this system. In this work, we theoretically study the structural, vibrational, and topological electronic properties of  Mo$_{0.5}$W$_{0.5}$Te$_2$ as a function of the hydrostatic pressure within 0 to 45 GPa  pressure range. We explore the aforementioned properties of two candidate crystal phases of Mo$_{0.5}$W$_{0.5}$Te$_2$, $T_d$ and 1$T''$, as a function of the applied pressure. Our calculations indicate that both these phases are dynamically stable within the studied pressure range and both of them host a number of WPs in their momentum-energy space. Interestingly, WPs in both phases can be created {\it via} pair-creation, destroyed {\it via} pair-annihilation, and shifted in the momentum-energy space by application of an external pressure. 
Our work implies that Mo$_{0.5}$W$_{0.5}$Te$_2$ inherits a variety of interesting topological properties at higher pressures and it may provide a novel platform for realization of Weyltronics. Furthermore, our results suggest an increase in the superconducting transition temperature at higher pressures, which is in agreement with recent experimental observations~\cite{TakahashiPRB2017,dahal2020tunable}.

\section{Computational Details}
The density-functional theory (DFT) calculations were performed using the Projector Augmented Wave (PAW) method as implemented in the Vienna Ab initio Simulation Package (VASP)~\cite{Kresse96a, KRESSE199615, KressePAW}. Six valence electrons were considered in the PAW pseudopotentials of Mo, W, and Te. The Perdew-Burke-Ernzerhof generalized-gradient approximation (GGA-PBE) was used to compute the exchange-correlation functional~\cite{PBEsol}. The GGA-opt86b functional was used for the van der Waals (vdW) density-functional corrections~\cite{PhysRevB.83.195131, PhysRevB.76.125112, PhysRevLett.92.246401}.
The reciprocal space was sampled using a $\Gamma$-centered k-mesh of size 8 $\times$ 12 $\times$ 4 together with a kinetic energy cutoff of 600\,eV for plane waves. 
The energy convergence and force convergence criteria were set to $10^{-8}$ eV and $10^{-4}$ eV/\AA, respectively. 
The hydrostatic pressure was applied up to 45 GPa. 
The crystal structures were fully optimized in inner coordinates as well in cell parameters for each pressure  considering spin-orbit coupling (SOC) and GGA-opt86b vdW corrections~\cite{PhysRevB.83.195131, PhysRevB.76.125112, PhysRevLett.92.246401}. All the DFT calculations were carried out in a twelve-atom unit cell of Mo$_{0.5}$W$_{0.5}$Te$_2$. 
The optimized lattice parameters for the $T_d$ phase at zero pressure are \textit{a} = 6.301, \textit{b} = 3.490, and \textit{c} = 14.076 \AA, and cell angles are $\alpha$ = $\beta$  =$\gamma$ = 90$^{\circ}$, which are in excellent agreement with the experimental data reported in Ref.~\cite{SchneelochPRB2020}.

The phonon calculations were performed using the finite-difference approach using supercells of size 2$\times$ 3 $\times$ 1. SOC was considered in all  phonon calculations. The {\sc phonopy}~\cite{TOGO20151} packaged was utilized to plot the phonon dispersions. The {\sc MechElastic}~\citep{mechelasticCPC2021,mechelasticPRB2018} package was used to perform the equation of states analyses using the enthaply versus pressure data. 
In order to understand the distribution of Weyl points in the momentum-energy space, we compute the real-space Wannier Hamiltonian using the full potential local orbital (FPLO) code~\cite{PhysRevB.59.1743}. We employed the above-mentioned DFT convergence parameter in all the FPLO calculations. 
The Wannier fitting was done using {\sc pyfplo} \cite{PhysRevB.59.1743} module of the FPLO package considering Mo:4\textit{d},\,5\textit{s}, W:5\textit{d},\,6\textit{s}, and Te:5\textit{s},\,5\textit{p}
as the local orbitals basis. The {\sc pyprocar} code~\cite{pyprocarCPC} was used to investigate the DFT calculated electronic band structures and  {\sc VESTA}~\cite{VESTA} software was used to draw the crystal structures.

\section{{RESULTS AND DISCUSSION}}

\subsection{Crystal structures}

\begin{figure*}[tbh]
\centerline{
\includegraphics[width = 6.75 in]{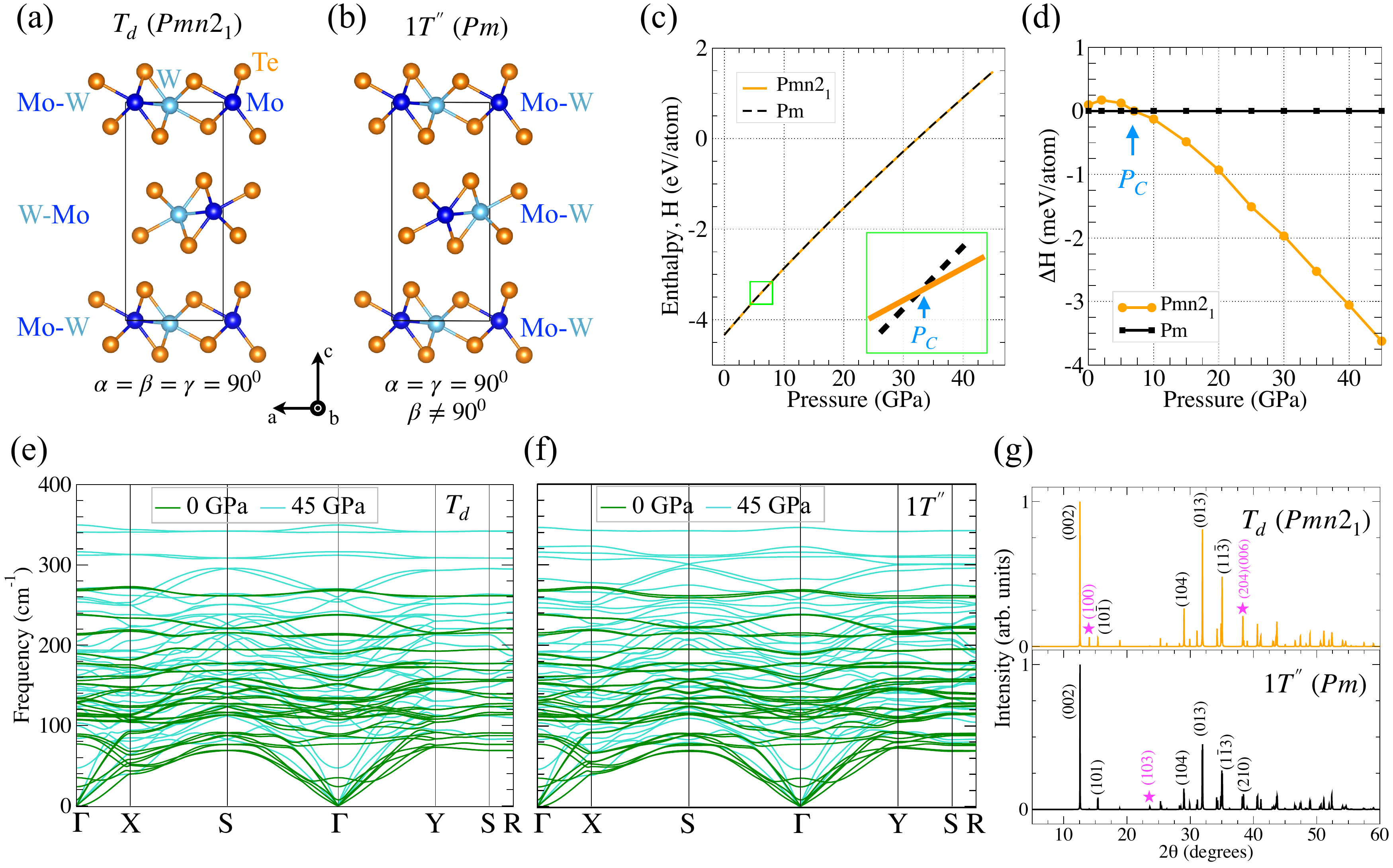}}
\caption{{
(a, b) Crystal structures of the $T_d$ and 1$T''$ phases of Mo$_{0.5}$W$_{0.5}$Te$_{2}$. Note that 1$T''$ has a subtle monoclinic distortion due to the cell angle $\beta \neq 90^{0}$. 
(c) Enthalpy versus pressure phase diagram for $T_d$ and 1$T''$ phases. The critical pressure ($P_C$) at which structural phase transition occurs is predicted near 7.5 GPa. Inset: an enlarged view near the crossing point. 
(d) The enthalpy difference $\Delta H$ (= $H_{T_d} - H_{1T''})$ as a function of pressure. 
(e, f) Calculated phonon spectra (with-SOC) at 0 and 45 GPa pressure for $T_d$ and 1$T''$ phases. 
We would like to note that we decided to plot the phonon dispersion for the 1$T''$ phase along the same high-symmetry k-path as in the $T_d$ phase for sake of better comparison, although 1$T''$ phase has a subtle monoclinic distortion (see SM~\cite{SM} for details).  
(g) The calculated x-ray diffraction (XRD) patterns for the $T_d$ and 1$T{''}$ phases at 0 GPa pressure. The signature peaks in both phases are marked using magenta color. The calculated XRD patterns and phonon spectra at higher pressures are reported in the SM~\cite{SM}.
}
}\label{fig:figure1}
\end{figure*}

In order to theoretically investigate the electronic structure of Mo$_{0.5}$W$_{0.5}$Te$_2$, we prepare two candidate crystal structures, as shown in Figs.~\ref{fig:figure1}(a,b), by taking the parent $T_d$ phase (space group $Pmn2_1$) of WTe$_2$ and systematically substituting one W atom by one Mo atom in each vertically-stacked layer. 
There are two possible ways to carry out such a substitution. 
First, we consider a scenario in which the ordering of Mo and W atoms is reversed in the adjacent vertically-stacked layers, i.e., (Mo-W)$\cdot\cdot\cdot$(W-Mo)$\cdot\cdot\cdot$(Mo-W)$\cdot\cdot\cdot$. This results in a crystal structure belonging to the space group $Pmn2_1$ (no.\,31), similar to the parent  $T_d$ phase, hence we call this structure the $T_d$ phase. Note, this structure retains the vertical glide-mirror symmetry, as discussed in Ref.~\cite{SinghPRL2020}. Also, this structure does not break the orthorhombic symmetry of the parent $T_d$ structure after a free DFT relaxation.

Second, we consider a scenario in which ordering of the Mo and W atoms remains the same within the adjacent vertically-stacked layers, i.e., (Mo-W)$\cdot\cdot\cdot$(Mo-W)$\cdot\cdot\cdot$(Mo-W)$\cdot\cdot\cdot$. Such a configuration breaks the vertical glide-mirror symmetry~\cite{SinghPRL2020} and results in the space group $Pm$ (no.\,6) after a free DFT relaxation of the unit cell. Note that the optimized structure is slightly distorted from the parent orthorhombic cell to a distorted monoclinic cell with a monoclinic cell angle of $\beta \neq 90^{\circ}$, which varies as a function of the applied hydrostatic pressure, as we discuss below. 
Although this monoclinic structure is similar to the 1$T'$ phase (space group: $P2_{1}/m$) of TMDs, it is lower in symmetry due to the broken inversion and vertical glide-mirror symmetries~\cite{SinghPRL2020}. Therefore, we decide to call this monoclinic phase as the 1$T''$ phase. We note both the structures, $T_d$ and 1$T''$, have broken inversion symmetry, which is a fundamental requirement for nonmagnetic Weyl semimetals.

Before moving further, let us briefly discuss the possible reason behind the observed monoclinic distortion in the 1$T''$ phase. 
One could imagine the presence of an in-plane polarity orientation determined by the peculiar ordering of Mo and W atoms having different electronegativity within each vertically-stacked layer, as shown in Figs.~\ref{fig:figure1}(a,b). In Fig.~\ref{fig:figure1}(a), where the ordering of Mo-W pairs is reversed as we move in the vertical direction, the in-plane polarity would reverse its sign in the adjacent vertically-stacked layers, i.e., $-+-+ \cdot\cdot\cdot$ or $+-+- \cdot\cdot\cdot$ (antipolar order). 
Here, $+/-$ sign denotes the polarity orientation parallel or antiparallel to the $a$ lattice vector. In such a configuration, a free relaxation of structure does not require any monoclinic distortion of the orthorhombic cell due to the perfect cancellation of dipolar-like interactions along the vertical direction (not strictly speaking, because partial screening of electric dipoles may occur due to the semimetallic nature of this system). 
On the other hand, in Fig.~\ref{fig:figure1}(b), where the ordering of Mo-W pairs is the same in the adjacent vertical layers, the in-plane polarity would be parallel in the adjacent layers, i.e., $---- \cdot\cdot\cdot$ or $++++ \cdot\cdot\cdot$, similar to that of the 1$T'$ phase~\cite{huang2019polar, SinghPRL2020}. Hence, in order to minimize the total free energy such structures tend to exhibit a monoclinic distortion due to the sliding of the adjacent polar layers along the in-plane direction, as discussed in Ref.~\cite{SinghPRL2020}.

\subsection{Pressure-induced effects on the $T_d$ and 1$T''$ structures} 

Next, we test the relative stability of the $T_d$ and 1$T''$ phases as a function of pressure. Our calculations reveal that the enthalpy difference ($\Delta H$) between these two phases is very small [Figs.~\ref{fig:figure1}(c, d)]. This implies the likelihood of formation of solid solution of the $T_d$ and 1$T''$ phases at finite temperatures. We find that below the critical pressure ($P_C$ = 7.5 GPa), the 1$T''$ phase is more favorable, whereas above $P_C$ the $T_d$ phase is preferred. 
We further performed the equation of states (EOS) analyses using the Birch, Vinet, and Birch-Murnaghan models as implemented in the {\sc mechelastic} package~\cite{mechelasticCPC2021}. 
The difference between the obtained EOS fitting parameters for the $T_d$ and 1$T''$ phases is minimal (see SM~\cite{SM} for more details). 

The phonon spectra calculated considering the SOC and vdW corrections for both the $T_d$ [Fig.~\ref{fig:figure1}(e)] and 1$T''$ [Fig.~\ref{fig:figure1}(f)] phases demonstrate the dynamical stability of these phases at zero pressure as well as at higher pressures (see SM~\cite{SM}). We notice an increase in the phonon frequencies with increasing pressure, i.e., phonons harden at higher pressures. No dynamical instability was observed in both phases within the studied pressure range. A list of the infrared- and Raman-active phonon frequencies calculated at different pressures is provided in the SM~\cite{SM}. We hope this could facilitate the experimental identification of the $T_d$ and 1$T''$ phases. 

Furthermore, we calculate the x-ray diffraction (XRD) patterns of the $T_d$ and 1$T''$ phases at various pressures using Cu-K$\alpha$ x-ray of wavelength 1.5406 \AA. Fig.~\ref{fig:figure1}(g) shows the calculated XRD patterns at zero pressure (data at higher pressure are provided in SM~\cite{SM}). Although the calculated XRD spectrum looks quite similar for both the phases, there are some signature peaks, marked using magenta stars, which are present in one phase but absent in another. These peaks can be used to distinguish between the $T_d$ and 1$T''$ phases in real  crystals.

\begin{figure}[tbh]
\centerline{
\includegraphics[width = 3.5 in]{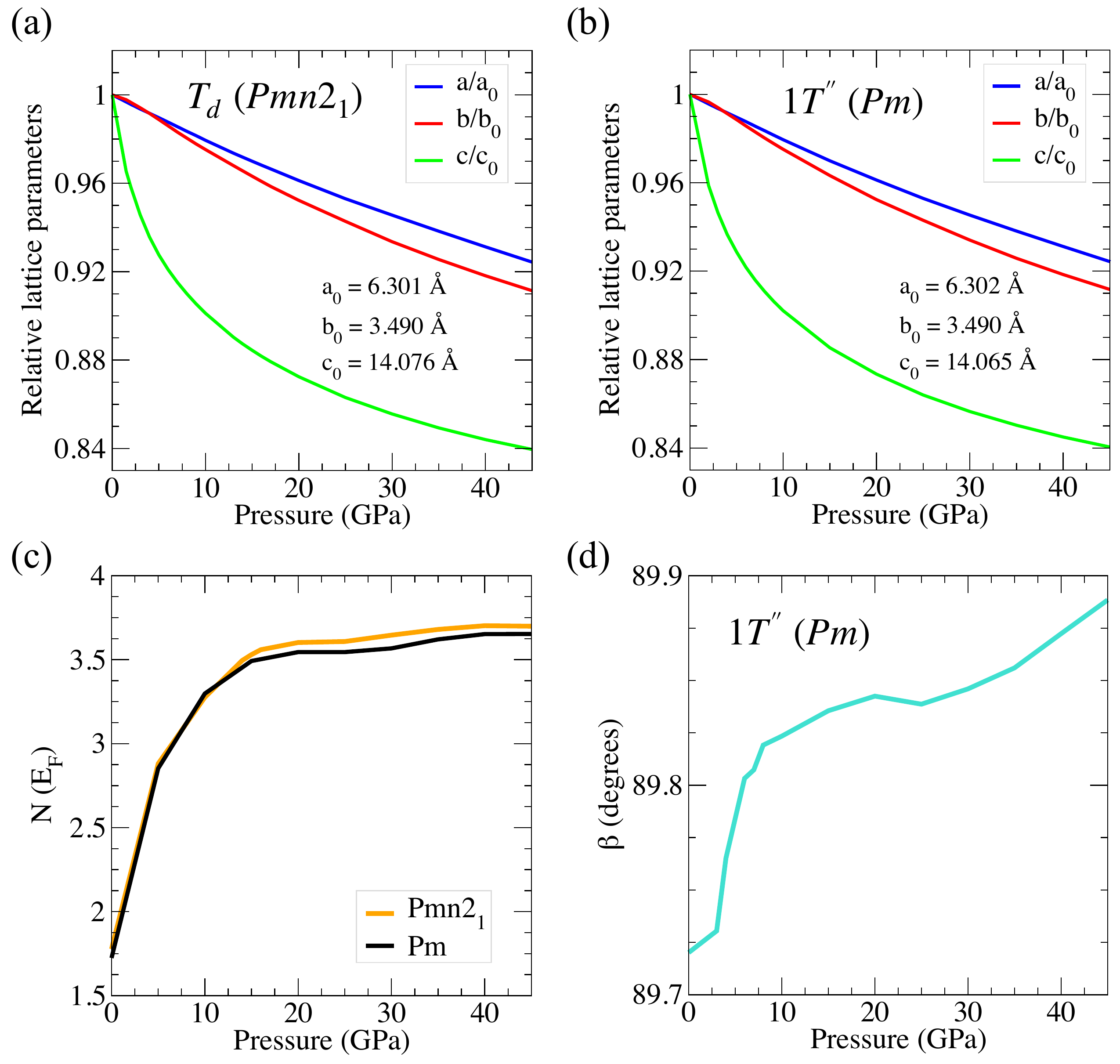}}
\caption{{
(a, b) The optimized lattice parameters normalized to the lattice parameters obtained at zero pressure ($a_0$, $b_0$, $c_0$), and (c) the calculated density of states (states/eV) at the Fermi level N(E$_{F}$) plotted as a function of pressure for both $T_d$ and 1$T''$ phases. 
(d) Pressure-dependent variation in the monoclinic cell angle ($\beta$) for 1$T''$ phase. Note that $\beta = 90^{0}$ for the $T_d$ phase. 
}
}\label{fig:figure2}
\end{figure}

Figs.~\ref{fig:figure2}(a,b) show the pressure dependence of the DFT optimized lattice parameters for $T_d$ and 1$T''$ phases. We observe a very similar trend in the pressure-dependent structural parameters of both phases. 
To highlight the observed trend, we plot the  normalized the lattice parameters with respect to the lattice parameters obtained at zero pressure. 
We observe maximum change in the $c$ lattice parameter with varying pressure. 
At 45 GPa, the $c$ lattice parameter decreases by $\sim$15\% for both phases, whereas relative change in the $a$ and $b$ lattice parameters is less than 10\% within the studied pressure range. Such a behavior is expected owing to the weak vdW interaction along the out-of-plane $c$ axis. 
The optimized Mo-W bond length shows a similar pressure-dependent behavior as the in-plane lattice parameters. The Mo-W bond length decreases from a value 2.85\,\AA~at zero pressure to 2.70\,\AA~at 45 GPa. This is almost 5\% decreases in the Mo-W bond length. On the other hand, the maximum compression in the Mo-Te and W-Te bond lengths is nearly 4\% at 45 GPa. Fig.~\ref{fig:figure2}(d) shows the variation in the monoclinic angle $\beta$ as a function of the pressure for the 1$T''$ phase. With increasing pressure, the cell angle $\beta$ tends to approach $90^{\circ}$, which implies a decrease in the monoclinic distortion and preference towards the orthorhombic $T_d$ phase at higher pressures. This is consistent with the data shown in Fig.~\ref{fig:figure1}(d).  

Since Mo$_{0.5}$W$_{0.5}$Te$_2$ is particularly interesting due to its superconducting properties~\cite{dahal2020tunable}, we calculate the density of states at the Fermi level N(E$_{F}$) as a function of pressure for both $T_d$ and 1$T''$ phases, as shown in Fig.~\ref{fig:figure2}(c). A systematic increase in N(E$_{F}$) with increasing pressure was observed in both phases. We would like to note that phonons also harden with increasing pressure [see Figs.~\ref{fig:figure1}(e,f)]. Thus, an increased N(E$_{F}$) together with the higher phonon frequencies implies an enhancement in the effective electron-phonon coupling at higher pressures, which could substantially increase the superconducting transition temperature in this system at higher pressures~\cite{DYNES1972615, LuPRB2016,HeikesPRM2018, PaudyalPRB2020}. 
This is consistent with recent experimental observations~\cite{dahal2020tunable, TakahashiPRB2017}.

\subsection{Pressure-tunable Weyl semimetal phase}

 \begin{figure*}[bth!]
 	 \includegraphics[scale=1.33]{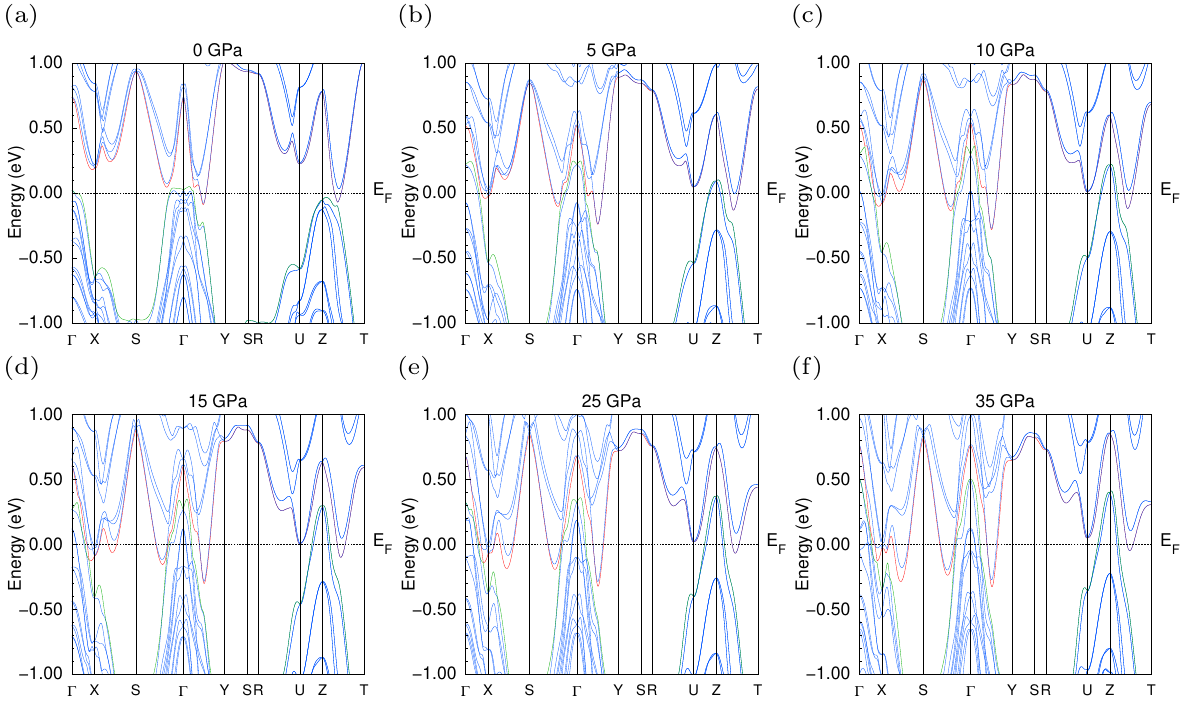}
 	\caption{Band structures of $T_d $ phase calculated at different pressure values. (a) 0 GPa, (b) 5 GPa, (c) 10 GPa and (d) 15 GPa, (e) 25 GPa, and (f) 35 GPa with inclusion of SOC. Red color in the band structure represents the lowest conduction band and the green color represents the highest valence band. The dashed horizontal line marks the E$_F$ (E$_F$ = 0 eV).}
 	\label{fig3}
 \end{figure*}
 
\begin{figure*}[bth!]
 	 \includegraphics[scale=1.33]{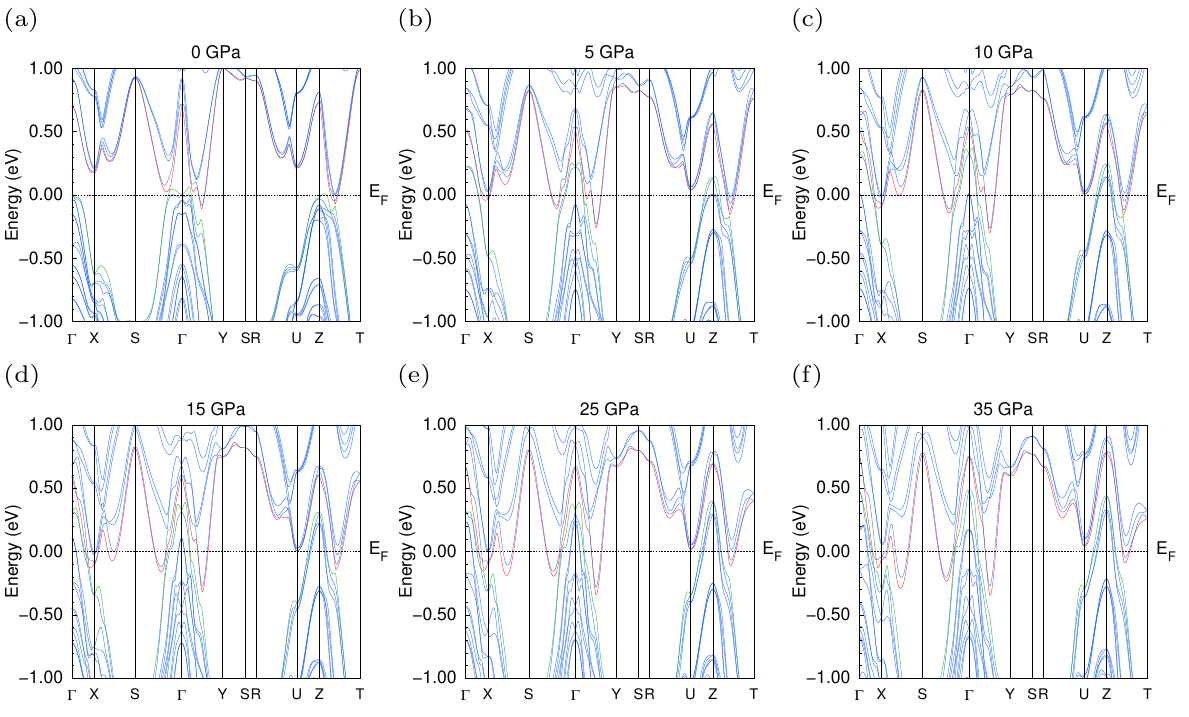}
 	\caption{Band structures of 1$T''$ phase calculated at different pressure values. (a) 0 GPa, (b) 5 GPa, (c) 10 GPa and (d) 15 GPa, (e) 25 GPa, and (f) 35 GPa with inclusion of SOC. Red color in the band structure represents the lowest conduction band and the green color represents the highest valence band. The dashed horizontal line marks the E$_F$. For the better comparison $k$ paths of 1$T''$ phase are also chosen same as in the $T_d$ phase.}
 	\label{fig4}
 \end{figure*}

\begin{figure*}[bth!]
  	 \includegraphics[scale=1.7]{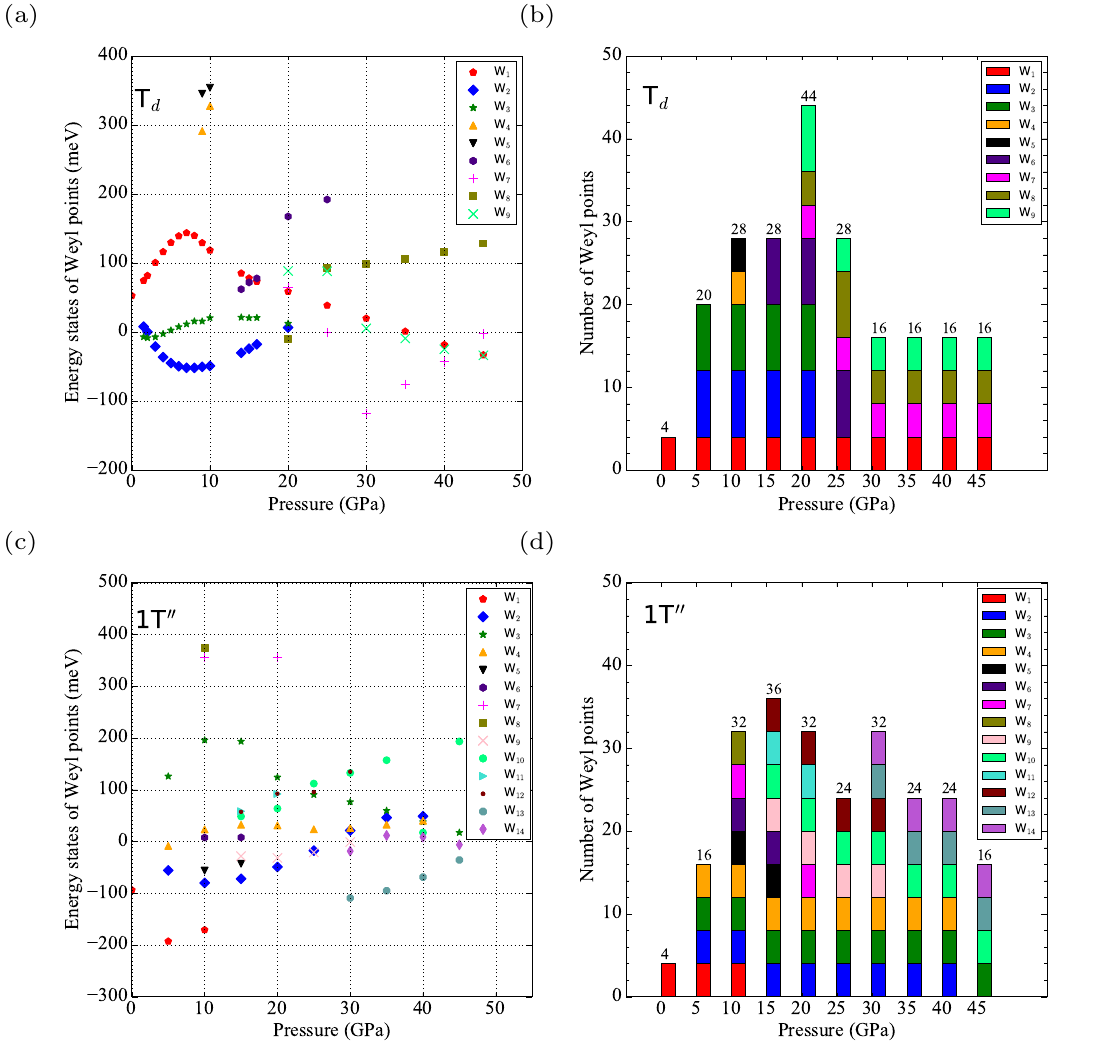}
  	\caption{(a) Energy states of WPs with respect to E$_F$ calculated at different values of pressure for the $T_d$ phase, (b) Stack-bar diagram showing the creation and annihilation of the WPs from 0 to 45 GPa for the $T_d$ phase. (c) Energy states of WPs corresponding to different pressure for the 1$T''$ phase. (d) Stack-bar diagram showing the creation and annihilation of the WPs from 0 to 45 GPa for the 1$T''$ phase. The total number of WPs at each pressure is written at the top of each bar.  }
 	\label{fig5}
 \end{figure*}

 The band structure calculations of $T_d$ and 1$T''$ phase for different pressure is presented in Figs.~\ref{fig3} and \ref{fig4}. The topmost valence band indicated with green color forms hole pocket and bottom of the conduction band indicated with red color forms electron pocket nearby the E$_F$ along the $\Gamma$-X, S-$\Gamma$-Y and U-Z-T high-symmetry directions of the Brillouin zone (BZ). These electron and the hole pockets touch each other at discrete points in the momentum space forming type-II WPs~\cite{soluyanov2015type}. The change in pressure affects the size of the electron and hole pockets in the momentum space. With increasing pressure, these pockets come close to each other producing more number of band crossings, as implied in Figs.~\ref{fig3} and \ref{fig4}. A similar trend is found for the 1$T''$ phase. \\

  TMDs are the first class of materials to host type II WPs. Soluyanov \textit{et al}. first reported type II WPs in  pure WTe$_2$  located between 0.052 -- 0.058 eV above E$_F$~\cite{soluyanov2015type}. 
  16 WPs are found in the $k$- space with 8 WPs at \textit{$k_z$} = 0 plane and 8 more WPs off the plane ($k_i$ $\neq$ 0, $i$ = $x$, $y$ and $z$). With inclusion of SOC all the WPs at $k_i$ $\neq$ 0 are annihilated leaving only eight of the sixteen WPs. Similarly, Sun \textit{et al}., predicted the type II WSMs in MoTe$_2$ \cite{sun2015prediction}. Eight WPs are recorded at two different  energies 6 meV and 59 meV above E$_F$.

 Here, in our study 50\% substitution of Mo on WTe$_2$ without application of pressure, we report four WPs at an energy state 55 meV above the E$_F$ at \textit{$k_z$} = 0 plane for the $T_d$ phase. 
 Contrary to that, a similar number of WPs are recorded for the 1$T''$ phase but at a different energy state of 93 meV below the E$_F$ at \textit{k$_z$} $\neq$ 0 plane. The energy of the WPs obtained for the $T_d$ phase is agreement with the energy of  WPs for MoTe$_2$~\cite{sun2015prediction} and WTe$_2$~\cite{soluyanov2015type}, whereas the 1$T''$ phase hosts WPs below E$_F$.
Moreover, the chemical effect is also found to play a significant role for the change in the total number of the WPs in Mo$_{0.5}$W$_{0.5}$Te$_2$.
\begin{table}[!h ]
  \centering
  \caption{Location of W$_1$ WPs in the momentum space of the $T_d$ and 1$T''$ phases at zero pressure}
  \begin{ruledtabular}
  \begin{tabular}{r c c c c }
WP &    $k_x$ $\left(\dfrac{2\pi}{a}\right)$    &  $k_y$ $\left(\dfrac{2\pi}{b}\right)$    &  $k_z$ $\left(\dfrac{2\pi}{c}\right)$ & Chirality($\chi$) \\\cline{1-5}
    
   W$_1$($T_d$) & -0.132 & -0.095 &  0.000&    +1\\
   W$_1$(1$T''$) & -0.004&   -0.200&   -0.416&  +1\\
  
  \end{tabular}
  \end{ruledtabular}
  \label{tab1}
 \end{table}
 
We obtain a total of four WPs (W$_1$) in the first BZ of the $T_d$ and 1$T''$ phases of Mo$_{0.5}$W$_{0.5}$Te$_2$ at zero pressure. Co-ordinates of the nonequivalent W$_1$ WPs are given in Table   \ref{tab1}; other three WPs are the mirror reflections of W$_1$ at zero
pressure. There is a variation in the total number and locations of the WPs due to the application of hydrostatic pressure. Below we discuss the role of the hydrostatic pressure on the Weyl phase of $T_d$- and $1T''$-Mo$_{0.5}$W$_{0.5}$Te$_2$. \\ 

\textbf{a) Pressure effects on the $T_d$ phase}

The variation in the number of the WPs and their energy states for the $T_d$ phase is presented through the graphical plots in Figs.~\ref{fig5} (a) and (b). On increasing the pressure value to 1.5 GPa the new sets of WPs W$_2$ and W$_3$ are generated along with the initial W$_1$ making the total number of WPs to 20. Here, we noticed that the pressure shifts the energy state of W$_1$ to E$_F$+75 meV. The newly created 8 copies of W$_2$ and W$_3$ are found in the energy states of E$_F$+8 meV and E$_F$-7 meV, respectively. No further creation of the WPs upto the pressure of 8 GPa, simply the position and the energy states of the WPs vary with the pressure. The energy state of W$_1$ rises to the highest value of E$_F$+139 meV at 8 GPa pressure and decreases on the further increase which reach nearest to the Fermi level (energy state E$_F$+1 meV) at 35 GPa and shifts below the Fermi level above that pressure. The pair creation of W$_4$ and W$_5$ each four in number occurs at 9 GPa which continue to exist up to 10 GPa and get annihilated above that pressure. The highest number of WPs is observed at 20 GPa and a constant number of 16 WPs is observed after 30 GPa to 45 GPa. \\

\textbf{b) Pressure effects on the 1$T''$ phase}

Pressure also has an impact on the dynamics of WPs in the 1$T''$ phase. Pair creation and the annihilation of WPs as the function of pressure is noticed in the momentum space of the 1$T''$ phase. 
Here, the W$_1$ WPs continue to exist up to 10 GPa which annihilates on further rise in pressure. 
Similarly, the W$_2$ WPs are created at 5 GPa pressure, which exist up to 40 GPa and then get annihilated. 
A similar phenomenon of pair creation and annihilation is noticed for the other energy state of WPs as well, which is summarized in the Figs.~\ref{fig5} (c) and (d). 
As compared to the $T_d$ phase more energy states of the WPs are observed in the 1$T''$ phase due to its lower symmetry. 
In terms of the total number of WPs, the highest number 36 is observed at 15 GPa pressure, whereas a constant number of 16 WPs is noticed for pressure beyond 40 GPa.

\section{Summary}
In summary, we studied the structural, vibrational, electronic, and topological Weyl properties of Mo$_{0.5}$W$_{0.5}$Te$_2$ by means of first-principles DFT calculations. We find that there are two possible candidate structures, $T_d$ and 1$T''$, for 50:50 Mo:W substitution. We studied the aforementioned properties of these two phases in 0-45 GPa pressure range. We find that both these structures are energetically and dynamically stable in the studied pressure. The calculated x-ray diffraction spectra, and the infrared- and Raman-active phonon frequencies indicate that these two phases can be identified in experiments, although they are likely to form solid solutions due to the subtle difference in their enthalpy at low pressures. The $T_d$ (1$T''$) phase is theoretically more favorable at higher (lower) pressures, the critical pressure being 7.5 GPa. Our calculations reveal that the density of states and phonon frequencies increase dramatically with increasing pressure in both phases, which is indicative of larger electron-phonon coupling at higher pressures and it could substantially increase the superconducting transition temperature in this system at higher pressures.

Interestingly, we find that both the $T_d$ and 1$T''$ phases host a number of WPs in their momentum-energy space. The total number and location of WPs can be controlled by varying hydrostatic pressure. Four WPs are obtained in the $T_d$ and 1$T''$ phases at zero pressure. The total number of WPs increases to 44 (36) with
increasing pressure, via pair creation, up to 20 (15) GPa for the $T_d$ (1$T''$) phase, and beyond this pressure, pair annihilation of WPs starts occurring leaving only 16 WPs at 45 GPa in both phases. Therefore, we can conclude that pressure can tune WPs to the desired location, generate new WPs, and also annihilate them, thus, providing an ideal platform for realization of Weyltronics in Mo$_{0.5}$W$_{0.5}$Te$_2$.

\acknowledgments
M.P.G. acknowledges the Alexander von Humboldt Foundation, Germany for the equipment grants and IFW-Dresden for providing the large-scale compute nodes to Advanced Materials Research Laboratory for scientific computations. S.S. acknowledges the support from the Office of Naval Research (ONR) grant N00014-21-1-2107. G.B.A thanks Nepal Academy of Science and Technology for the PhD fellowship. M.P.G. and G.B.A thanks Manuel Richter for the fruitful discussion and Ulrike Nitzsche for the technical assistance. 
   \pagebreak
\bibliography{ref}

\end{document}